ARTICLE

# SynTwins: A Retrosynthesis-Guided Framework for Synthesizable Molecular Analog Generation

Shuan Chen,[ab] Gunwook Nam[a] , Alan Aspuru-Guzik[c], and Yousung Jung*[abd]

The disconnect between AI-generated molecules with desirable properties and their synthetic feasibility remains a critical bottleneck in computational discovery of drugs and materials. While generative AI has accelerated the proposal of candidate molecules, many of these structures prove challenging or impossible to synthesize using established chemical reactions. Here, we introduce SynTwins, a novel retrosynthesis-guided molecule design framework that finds synthetically accessible molecular analogs by emulating expert chemists' strategies in three steps: retrosynthesis, searching similar building blocks, and virtual synthesis. Using a search algorithm instead of a stochastic data-driven generator, SynTwins outperforms state-of-the-art machine learning models at exploring synthetically accessible analogs while maintaining high structural similarity to original target molecules. Furthermore, when integrated into existing molecular property-optimization frameworks, our hybrid approach produces synthetically feasible analogs with minimal loss in property scores. Our comprehensive benchmarking across diverse molecular datasets demonstrates that SynTwins effectively bridges the gap between computational design and experimental synthesis, providing a practical solution for accelerating the discovery of synthesizable molecules with desired properties for a wide range of applications.

[a] *Department of Chemical and Biological Engineering, and Institute of Chemical Processes, Seoul National University, 1 Gwanak-ro, Seoul, South Korea.*
[b] *Institute of Engineering Research, Seoul National University, 1 Gwanak-ro, Seoul, South Korea*
[c] Department of Chemistry, Department of Computer Science, University of Toronto, Vector Institute for Artificial Intelligence, Canadian Institute for Advanced Research, Toronto, Ontario, Canada
[d] *Interdisciplinary Program in Artificial Intelligence, Seoul National University, 1 Gwanak-ro, Seoul, South Korea. E-mail: yousung.jung@snu.ac.kr*

† Electronic supplementary information (ESI) available: Methods and materials, ablation study, and the additional details of the experimental results. See DOI: See DOI: 10.1039/x0xx00000x

## Introduction

The discovery of novel molecules with specific chemical properties is a critical yet challenging process in pharmaceutical and chemical industries, often requiring years or decades due to the vastness of chemical space[1,2]. Generative artificial intelligence (AI) has emerged as a powerful accelerator for this process, rapidly proposing candidate molecules with target properties[3,4]. When paired with accurate property prediction models, these AI approaches enable efficient computational screening of molecules against desired criteria. However, a fundamental limitation persists: a significant portion of AI-generated molecules are difficult or impossible to synthesize using known chemical reactions and available building blocks[5]. While synthetic accessibility scoring can partially guide generative models toward more feasible structures, many proposed molecules remain synthetically challenging, creating a disconnect between computational design and experimental implementation.

To address this synthesis planning bottleneck, computational tools for single-step and multi-step retrosynthesis have been developed[6–8]. However, even with these tools, many AI-generated molecules remain synthetically challenging. In such cases, generating structurally similar synthetically accessible analogs offers a promising alternative to bypass synthesis difficulties while preserving desired properties, similar to how medicinal chemists have designed accessible analogs of complex natural products[9,10]. In real-world laboratory settings, synthesis capabilities vary considerably—pharmaceutical labs frequently rely on amide formations and heterocycle synthesis, while materials science teams favour reactions like Suzuki coupling for developing OLEDs. Additionally, the growing emphasis on sustainable chemistry has shifted preferences toward greener reactions and building blocks[11–13].

Given these constraints, the set of synthetically accessible molecules varies across different research environments. Consequently, researchers are exploring strategies for designing molecules by virtually synthesizing them from predefined reaction sets and available building block libraries. For example, the virtual on-demand libraries constructed by computationally enumerating in-house reactions and compounds from different pharmaceutical companies and chemical suppliers were reported to have the collections of more than 1 billion molecules[14–16]. On the other hand, Levin et al.[17] developed an algorithm to explore diverse and synthetically accessible analogs by enumerating different building blocks to replace the precursors in a known synthesis pathway of one molecule.

Recent advances have applied machine learning (ML) to address these challenges through two main strategies. The first approach focuses on explicitly generating synthesizable molecules by selecting reactants and reaction pathways optimized for both synthetic feasibility and target properties using machine learning models[18–21] (**Fig. 1b**). The second strategy designs synthetically accessible analogs, structurally similar alternatives to promising but challenging-to-synthesize AI-generated molecules (**Fig. 1c**). For example, Noh et al.[22] created a variational autoencoder (VAE) that encodes multi-step synthesis sequences into a latent space to generate complex yet synthetically feasible molecules. Similarly, Gao et al.[23] formulated the problem as a Markov decision process (MDP) with an amortized solution approach. While these methods represent promising directions, existing models typically achieve suboptimal structural similarity to target molecules, limiting their practical utility.

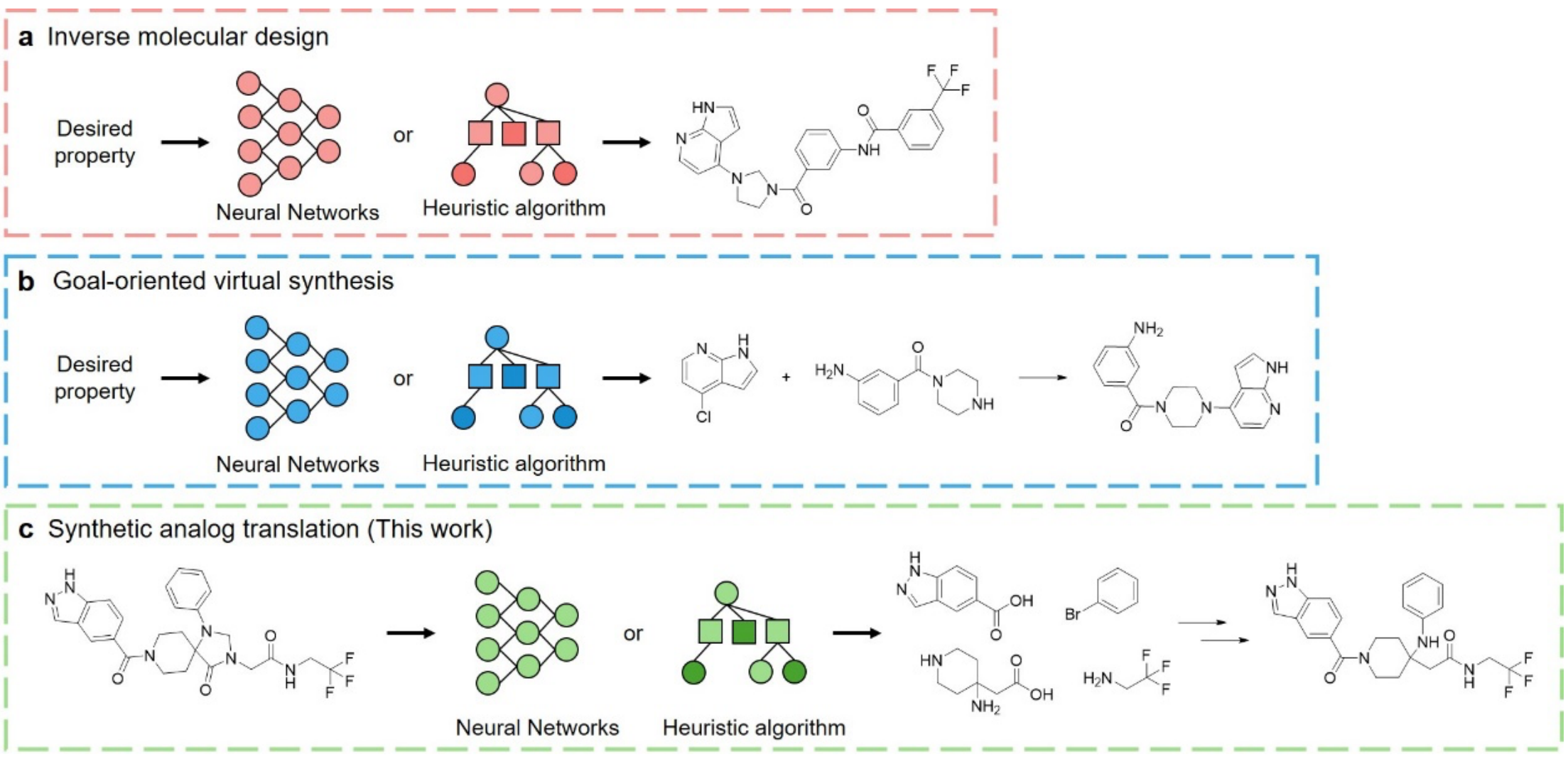

**Fig. 1.** The three strategies for designing novel molecules using computational methods. (a) Inverse molecular design model. (b) Conditional virtual synthesis model. (c) Synthetically accessible analog design.

We introduce SynTwins, a novel tree search algorithm that designs synthetically accessible molecular analogs by emulating the intuitive strategies employed by experienced medicinal and synthetic chemists. SynTwins implements a three-step process: (1) retrosynthetic analysis of target molecules to identify key structural components, (2) systematic searching for similar yet readily available building blocks that maintain essential pharmacophores, and (3) virtual synthesis using well-established reaction templates to assemble the final molecular analog. While the molecule analog generation process is similar to the approach described in Levin et al.[17], instead of diversifying, SynTwins generates a set of structurally similar and synthetically accessible molecule analogs from a reference molecule. Our comprehensive evaluation demonstrates that SynTwins, a non-ML-based approach, outperforms state-of-the-art ML-based models in generating synthetically accessible analogs preserving high structural similarity to original target molecules. SynTwins prioritizes generating synthetically accessible analogs with high structural similarity to enable systematic property optimization. Importantly, when complete retrosynthetic reconstruction fails, SynTwins leverages partial decompositions to discover alternative routes using available chemistry components.

Furthermore, by integrating SynTwins with existing ML-based molecule optimization pipelines, our hybrid approach produces synthetically feasible molecules with comparable bioactivity and physicochemical profiles to those generated by unconstrained molecular optimizers. This work presents a practical solution to the synthesis-design gap and establishes a foundation for more effective molecular discovery pipelines that successfully translate computational designs into laboratory syntheses.

## Results and Discussion

### SynTwins

*SynTwins* is a synthetically accessible analog search algorithm that mirrors the intuitive workflow of chemists in a laboratory. Typically, chemists first evaluate whether a target molecule can be synthesized directly using available reactions and building blocks. If direct synthesis is not feasible due to the unavailability of specific precursors, they explore alternative building blocks with structural similarities to construct analogous molecule. Following this intuition, SynTwins systematically explores potential analogs of a target molecule through a combined iterative process of retrosynthesis analysis and virtual synthesis. The algorithm operates in three distinct phases: retrosynthesis, building block search, and virtual synthesis. Computationally, the full processes are achieved by the following three phases of SynTwins (**Fig. 2**):

1. Retrosynthesis: The first step of SynTwins is to perform multi-step retrosynthesis using retro-reaction templates[24] (denoted as $T_r$, detailed in the next subsection) to build a synthesis tree. For retrosynthesis, we apply all the compatible retro-reaction templates to the target molecule at each retrosynthesis step instead of applying ML-based retrosynthesis models in this work. The synthesis tree is expanded until maximum tree depth ($d_{max}$) is reached. The default number of $d_{max}$ is 3 in this study. The target molecule and the precursors found in the synthesis tree are denoted as $P_{ref}$ and $\{R_{ref}\}$, where $\{R_{ref}\}$ may or may not include the ones in the set of available building blocks. More details about the retrosynthesis are given in the **Electronic Supplementary Information (ESI)†**.
2. Building block search: After the retrosynthesis is finished, the building blocks that are structurally similar to the found precursors $\{R_{ref}\}$ are searched using an $k$-nearest neighbor (kNN) algorithm[25], which search for the molecules that have the shortest distance in the chemical space according to their Extended Connectivity Fingerprints (ECFP)[26]. The building blocks found by the kNN algorithms are denoted as $\{R_{twin}\}$, and these molecules are limited to have the same functional groups presented in $T_r$ to guarantee the compatibility of applying the reverse reaction templates used for the virtual synthesis in the next step. The default number of neighbors ($k$) of kNN search is 10 in this study. More details about the kNN implementation are given in the **Electronic Supplementary Information (ESI)†**.
3. Virtual synthesis: Finally, $k^2$ molecule analogs $\{P_{twin}\}$ are synthesized by applying the forward-reaction template ($T_f$), which represent the reverse version of the retro-reaction template $T_r$ used in the first phase, on every pair of building block searched in the last phase $\{R_{twin}\}$.

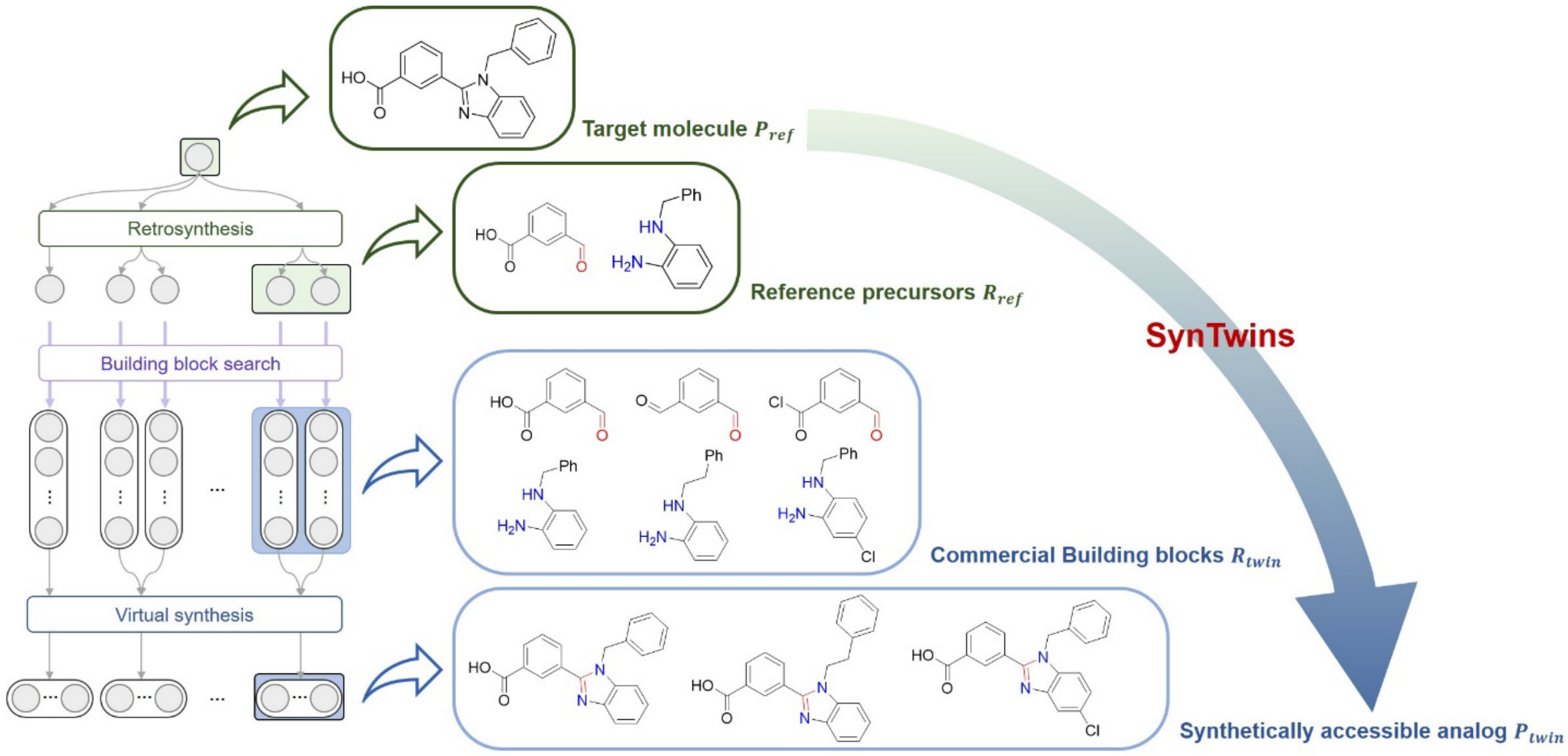

**Fig. 2.** An example of generating synthetically accessible analogs using SynTwins. The imidazole in the target molecule $P_{ref}$ is decoupled to two precursors $R_{ref}$ containing an aldehyde and diamine functional groups, respectively. Next, the building blocks structurally similar to the precursors $R_{twin}$ were searched from the list of available molecules. Finally, the synthetically accessible analogs $P_{twin}$ are virtually synthesized by the same imidazole synthesis reaction using the aldehyde and diamine functional groups in the building blocks.

**Retro-reaction templates**

The chemical reactions available in this study are represented by forward-reaction templates in the form of Simplified Molecular Input Line Entry System of Reactions (SMIRKS), where the reactive substructure patterns are described in the SMILES arbitrary target specification (SMARTS) format[27]. To derive retro-reaction templates from the known forward-reaction templates, we virtually synthesized multi-step (from 3 to 5 steps) chemical reactions by applying the forward-reaction templates to the available building blocks and extracted the retro-reaction templates using a modified version of RDChiral[28]. More details are given in the **Electronic Supplementary Information (ESI)†**.

To balance the specificity and generalizability of retro-reaction templates, the extracted retro-reaction template includes the reaction center and its one-hop neighboring atoms, without explicitly defining functional groups. This representation is similar to the extended local reaction template (ELRT) used in LocalMapper[29]. Consequently, the number of retro-reaction templates is approximately 10 times greater than the number of forward-reaction templates in our experiments.

**Synthesizable accessible analog design**

To run SynTwins for our experiments, we followed existing works[18,22,23,30,31] and collected 58 reaction templates from Hartenfeller et al.[32] and 64 reaction templates from Button et al.[33]. After removing the 21 duplicated reactions, where different reaction templates from different literatures represent the same chemical reaction, 101 forward-reaction templates were remained from the total 122 forward-reaction templates.

We use the same set of 150,560 commercially available molecules from the Enamine Building Block catalogue Global Stock[34] used in the previous molecule analog generation studies[22,35]. Using these 101 forward-reaction templates and 150,560 building blocks, we virtually synthesized 1.01 million reactions and extracted 18,590 retro-reaction templates according to their atom-mappings. After filtering the retro-reaction templates that appear less than 100 times in the 1.01 million reactions, we obtained 1,163 retro-reaction templates as our final retro-reaction template collections. More details about the selected reaction templates are given in the **Electronic Supplementary Information (ESI)†.**

To evaluate the performance of SynTwins on generating the synthetically accessible analogs, we compare our method with two state-of-the-art ML-based models, ChemProjector[30] and SynFormer[31], after training the models with the same set of reaction templates and building blocks. We note that these methods employ fundamentally different strategies: SynTwins uses retrosynthesis-guided building block search, whereas ML models generate analogs through learned representations. To ensure fair comparison, we provide all methods with identical chemical resources: the same 101 forward-reaction templates and 150,560 Enamine building blocks. While SynTwins uses retro-reaction templates for its framework, ChemProjector and SynFormer's forward-only architectures cannot integrate retro-reaction templates without a complete redesign, as they generate analogs directly without retrosynthetic decomposition. More details of training the baseline models are given in the **Electronic Supplementary Information (ESI)†**.

We examine the performance of these algorithms using 1,000 virtually synthesized products, 1,000 molecules from ChEMBL[36], 170 molecules from US Patent Trade Office (USPTO) dataset[37], and 100 molecules from FDA-approved drugs[38]. The details of curating test datasets are given in the **Electronic Supplementary Information (ESI)†**.

The results are evaluated by exact-match rate and top-k average similarity. The exact-match rate shows the percentage of generated molecule analogs being exactly same with the target molecules, and the top-k average similarity is a metric to measure how structurally similar are the top-k generated molecule analogs are compared to the target molecule. The structural similarity is calculated by the Tanimoto similarity[39,40] of 4096-bits ECFP4[26] .

As shown in **Table 1**, SynTwins consistently outperforms the baseline methods across all datasets in both exact-match capability and molecular similarity. Compared to ChemProjector and SynFormer, SynTwins ranks first in exact-match performance, especially notable in realistic molecules like USPTO molecules and FDA-approved drugs, where other methods struggle to recover most of the original molecules. This suggests that SynTwins is better designed to handle both virtual and real-world molecular structures. Notably, although the approximation process of SynTwins inevitably sacrifices a significant portion of the exact-match rate, it can generate synthetically accessible analogs that are highly structurally similar to the targets.

**Table 1.** The exact-match rate and the top-k average similarity of generated molecules on the benchmarked datasets. The highest values are highlighted in bold fonts.

| Test data | Model | Exact-match rate | Top-k average similarity | | |
|---|---|---|---|---|---|
| | | | k = 1 | 3 | 5 |
| Virtual molecules | ChemProjector | 39.8% | 0.8018 | 0.7554 | 0.7268 |
| | SynFormer | 7% | 0.6543 | 0.6247 | 0.6070 |
| | SynTwins (this work) | **55%** | **0.8701** | **0.8209** | **0.7992** |
| ChEMBL molecules | ChemProjector | 10.8% | 0.5612 | 0.5256 | 0.5043 |
| | SynFormer | 8.9% | 0.5725 | 0.5397 | 0.5221 |
| | SynTwins (this work) | **19.6%** | **0.6630** | **0.6222** | **0.6025** |
| USPTO molecules | ChemProjector | 0% | 0.4225 | 0.4103 | 0.4009 |
| | SynFormer | 0% | 0.4323 | 0.4204 | 0.4114 |
| | SynTwins (this work) | **1.8%** | **0.5298** | **0.5064** | **0.4941** |
| FDA-approved drugs | ChemProjector | 1% | 0.4543 | 0.4382 | 0.4258 |
| | SynFormer | 4% | 0.3294 | 0.3089 | 0.2979 |
| | SynTwins (this work) | **17%** | **0.6387** | **0.6051** | **0.5873** |

In terms of structural similarity, SynTwins also leads across all datasets, producing molecules that are more structurally similar to the target chemicals even when exact-match is not achieved. While ChemProjector performs moderately well, especially with virtual molecules, it falls short in real-world cases. SynFormer, though competitive for the ChEMBL test set, consistently ranks lowest overall. These results highlight the versatility and robustness of SynTwins in generating high-quality molecules across diverse chemical spaces. The ablation study of using different depth of the synthesis tree, number of neighbors, size of the fingerprint vector, and fingerprint radius used in SynTwins can be found in the **Electronic Supplementary Information (ESI)†**.

Next, we analyze the relationship between the heuristic synthetic accessibility score and the top-5 similarity of molecule analogs generated by SynTwins for the USPTO molecules in **Fig. 3**. The synthetic accessibility score of a molecule is calculated by the building block and reaction-aware SAScore (BR-SAScore)[41] using the reactions and building blocks available in this study. For reading clarity, we show BR-SAScores and molecular similarities of 10 molecules with the lowest BR-SAScores (easy-to-synthesize) and 10 molecules with the highest BR-SAScores (hard-to-synthesize) sampled from the USPTO molecules in **Fig. 3a** and **3b**. Similar plots for the FDA-approved drugs, and comparison with ChemProjector and SynFormer for all USPTO and FDA-approved molecules are provided in the **Electronic Supplementary Information (ESI)†**.

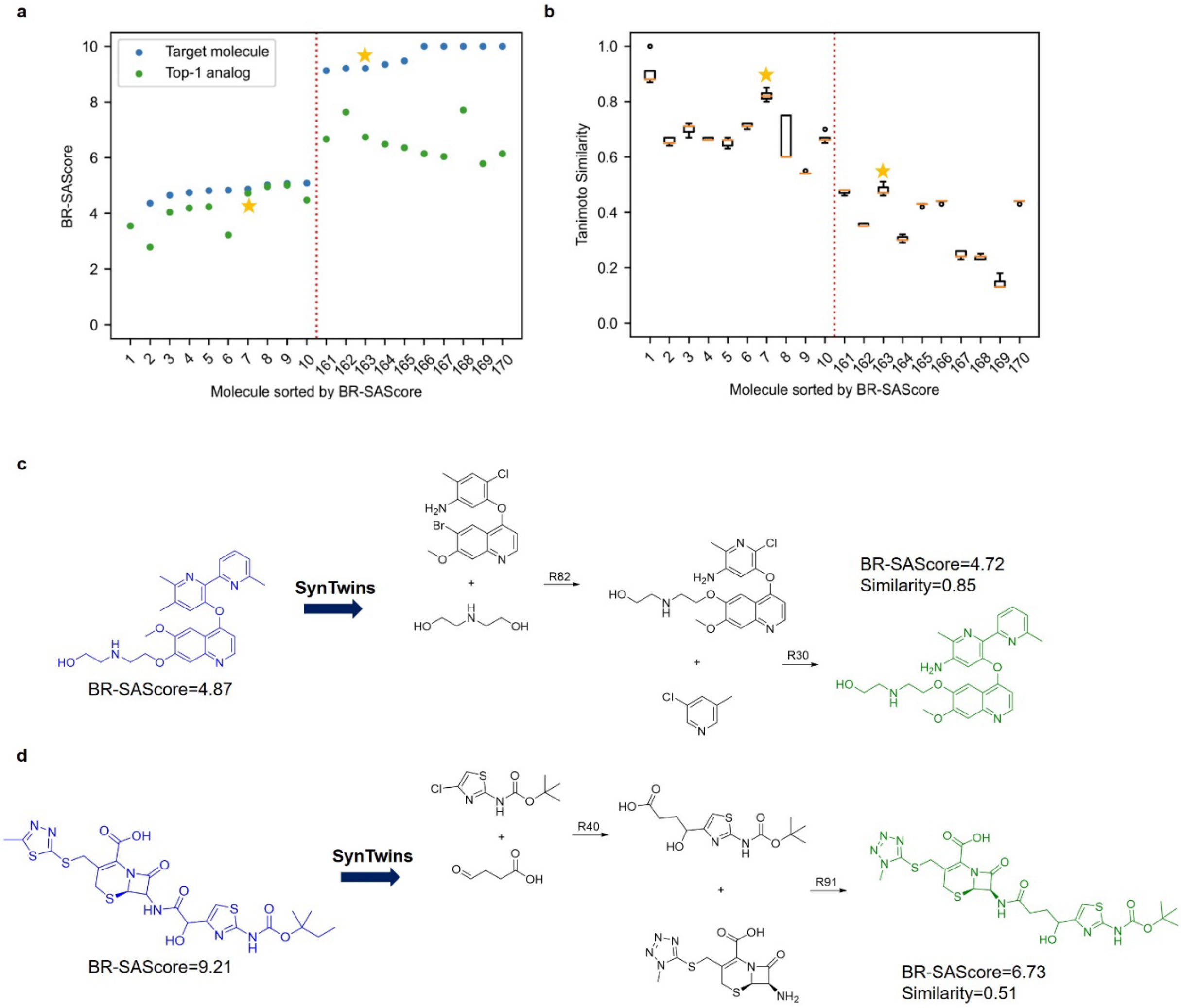

**Fig. 3.** The statistics and examples of synthetically accessible analogs generated by SynTwins from the molecules sampled from the USPTO dataset[37]. The example molecules in (c) and (d) are highlighted by star symbols in (a) and (b). (a) The BR-SAScores of the target molecules and the most similar molecular analogs generated by SynTwins. (b) The box plots of the top-5 similarity of the molecular analogs generated by SynTwins. (c) An example of the most similar molecular analog generated from an easy-to-synthesize target molecule by SynTwins. The target molecule is colored in blue, and the generated molecular analog is colored in green. (d) An example of the most similar molecular analog generated from a hard-to-synthesize target molecule by SynTwins. The target molecule is colored in blue, and the generated molecular analog is colored in green. R82: Williamson reaction, R30: Negishi coupling, R40: Grignard reaction, R91: Amide formation. The full list of the reaction can be found in the **Electronic Supplementary Information (ESI)†**.

As expected, molecules with lower BR-SAScores are easier for SynTwins to search the synthetically accessible analogs with higher similarity. Notably, the 7 molecules (1 from USPTO-190 molecules and 6 from FDA-approved drugs) having BR-SAScore lower than 4 are all successfully reproduced by SynTwins. In contrast, SynTwins struggles to generate molecules having high BR-SAScores, resulting molecule analogs with low structural similarities. Furthermore, the generated molecular analogs consistently exhibited lower BR-SAScores than their corresponding target molecules. This decrease was more evident for hard-to-synthesize molecules than for easy-to-synthesize ones.

To visualize the how the Tanimoto similarity aligns with the chemist's view of structural similarity, we show the target molecules and the analogs generated by SynTwins for one easy-to-synthesize and one hard-to-synthesize molecules sampled from the USPTO molecules in **Fig. 3c** and **3d**. For easy-to-synthesize molecules (**Fig. 3c**), SynTwins is able to generate a highly structurally similar molecule analog with a 0.85 similarity score. On the other hand, for hard-to-synthesize molecules (**Fig. 3d**), SynTwins struggles to generate structurally similar molecules with a 0.51 similarity score. We note that even if the generated molecules are structurally different from the original molecules, especially for the complex molecular structures, their bioactivity can be similar when the crucial functional groups preserved. For instance, the beta-lactam substructure, a well-known functional group for antibiotics, in the target shown in **Fig. 3c** is preserved in the generated synthetically accessible analogs despite the low structural similarity.

To understand instances where SynTwins failed to generate exact molecular matches from reference compounds, we analyzed the building blocks and reactions employed in the USPTO dataset and assessed whether SynTwins could reproduce the reported synthetic routes. Our analysis revealed that none of the synthetic pathways used to prepare the 170 target molecules could be fully reproduced using the building block and reaction sets collected in this study. Of the 1,268 reaction steps comprising the 170 reported synthetic routes, a significant proportion of the required building blocks were unavailable (48.9%), and an even larger fraction of reactions were absent from our collected reaction set (84.2%). Nonetheless, SynTwins successfully proposed alternative synthetic pathways for 3 of the 170 USPTO molecules, yielding a 1.8% exact-match rate. Additional analysis details and examples are provided in the **Electronic Supplementary Information (ESI)†**.

### Molecule optimization

Here, we explore the application of embedding SynTwins into the existing molecule optimization methods to generate optimized molecules that are synthetically accessible by the available reactions and building blocks. Here, we selected the two leading algorithms for molecule optimization, REINVENT[42] and GraphGA[43], according to the practical molecule optimization (PMO) benchmark[44]. For each algorithm, we designed a synthesizable molecule optimizer variant that converts each optimized molecule into a molecule analog using SynTwins during the optimization process, which guarantees the synthetic accessibility of all the generated molecules (**Fig. 4a**). For comparison, we also compare the results by using SynTwins to convert the molecules optimized at the end of the optimization process. The results of the former methods are denoted as "Syn-REINVENT" and "Syn-GraphGA", and the later methods are denoted as REINVENT* and GraphGA* for the variants using REINVENT and GraphGA as molecule optimization backends, respectively. We perform the optimization on 7 similarity-related multi-property objective (MPO) tasks selected from GuacaMol[45]. We only use the top-1 molecule analog generated from SynTwins with $d = 2$ and $k = 1$ in this experiment. The scoring criteria of each MPO task is given in the **Electronic Supplementary Information (ESI)†**.

The top-10 scores of the molecules generated by the comparing algorithms for the 7 MPO tasks are shown in **Fig. 4b**. Overall, unconstrained REINVENT and GraphGA shows higher top-10 score and higher BR-SAScores than their variants. Nonetheless, these (potentially unsynthesizable) optimized molecules exhibit huge MPO score drops after they were converted to synthetically accessible analogs by SynTwins, with 0.351 and 0.498 difference for REINVENT and GraphGA, respectively. On the other hand, the proposed synthesizable molecule optimizers, Syn-REINVENT and Syn-GraphGA, show slightly lower MPO performance than regular molecule optimizers while guaranteeing their synthetic accessibility and low BR-SAScores. The top-1 and top-100 scores of the 7 MPO tasks are provided in the **Electronic Supplementary Information (ESI)†**.

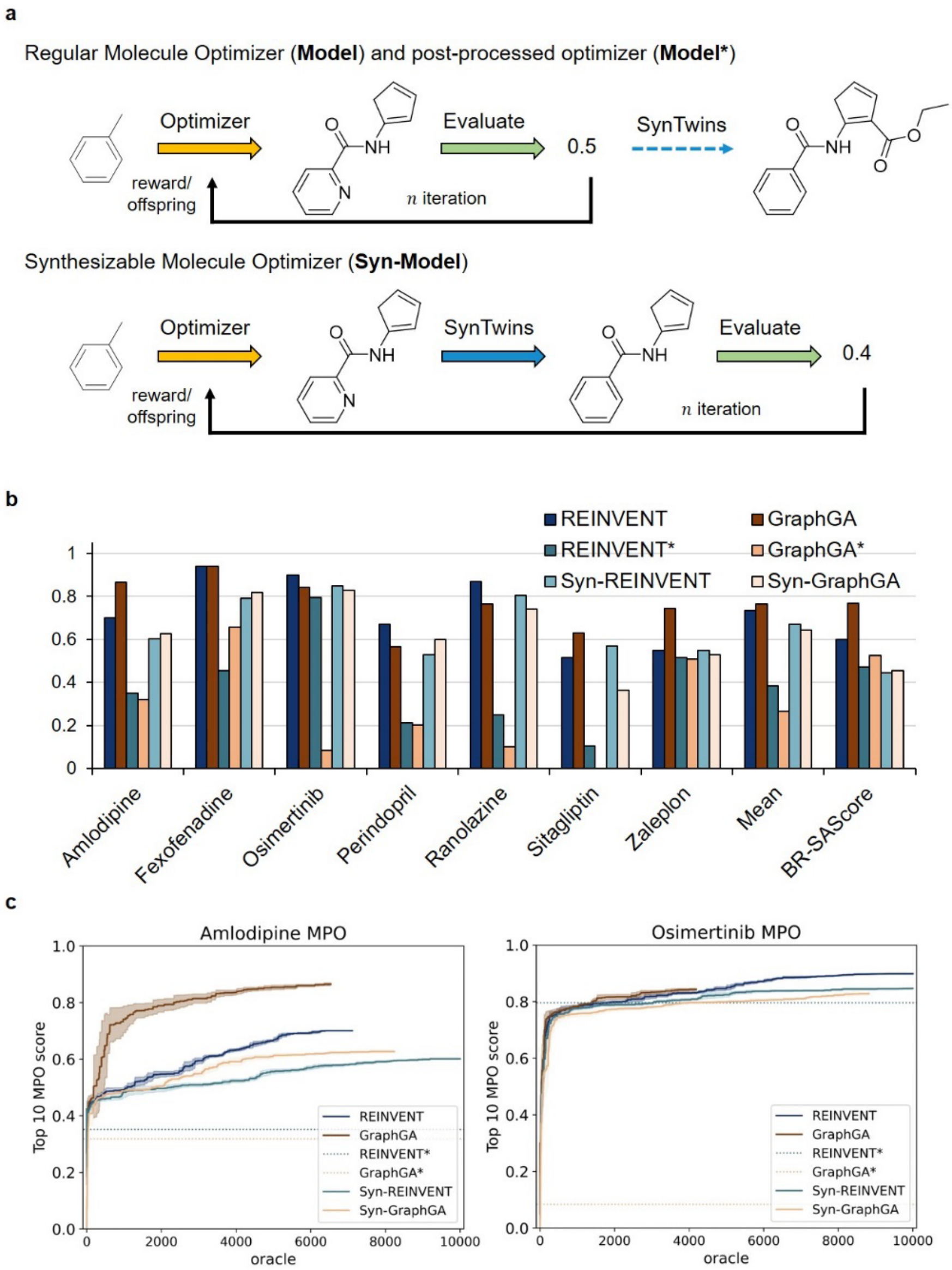

**Fig. 4.** The workflow and results of embedding SynTwins in the molecule optimization tasks. (a) The comparison of the regular molecule optimizer (Model) and the synthesizable molecule optimizer (Syn-Model). SynTwins can be applied to the final molecule generated by the regular molecule optimizer to convert the optimized molecules into synthetically accessible molecules (Model*). (b) The top-10 scores on 7 multi-property optimization (MPO) tasks from GuacaMol[45] and the BR-SAScores of the compared models. The BR-SAScores are rescaled by a factor of 0.1 to match the scale of other metrics. (c) The optimization process of top-10 scores of compared molecule optimizers on two MPO tasks. The concrete lines represent the top-10 average scores and the shadow near lines are the standard deviations.

We analyze the optimization curves of the optimizers and show two of them in **Fig. 4c**. We observed that integrating SynTwins into the optimization algorithms leads to longer converging time than the original algorithms, showing the increased difficulty of optimizing the molecules in a synthetically accessible molecule space. Particularly, we found several molecules generated by the original optimization algorithms collapse to the same molecules during the analog searching process, making the time of optimizing the synthetically accessible molecules significantly longer than that of the regular optimization algorithms. The optimization plots of all the 7 tasks for top-1, top-10 and top-100 molecules can be found in the **Electronic Supplementary Information (ESI)†**.

## Conclusions

We presented SynTwins as a robust and powerful tool for designing synthetically accessible molecule analogs for target molecules. The proposed retrosynthesis-guided molecular analog design framework leverages both retro-reaction and forward-reaction templates to design analogs within a limited set of reactions and building blocks. The key advantages of SynTwins compared to existing methods are twofold. First, unlike previous bottom-up approaches[22,23,30,31,33] that directly sample building blocks based on an arbitrary embedding of the target molecule, SynTwins employs a top-down precursor-searching strategy that aligns with chemists' practice of designing molecular analogs using available building blocks. By mimicking this intuition through a three-step process (retrosynthesis, similar building block searching, and virtual synthesis), SynTwins provides interpretable results and allows for easy optimization to better reflect chemists' needs. Second, SynTwins does not rely on machine learning models, making it more robust to variations in training hyperparameters and hardware constraints. For instance, SynFormer requires over 1,000 GPU hours for training[31], and retraining whenever new chemistry becomes available. SynTwins avoids this by instantly incorporating new building blocks or reactions on standard computers, making it practical for iterative workflows. Moreover, it can flexibly adapt to different reaction conditions and building block sets without requiring retraining from scratch. SynTwins has the potential to be embedded into molecular optimization workflows, enabling synthesis-aware molecular design.

We acknowledge that SynTwins and ML-based methods represent fundamentally different approaches with distinct trade-offs. SynTwins' performance depends on finding valid synthetic pathways within the provided reaction templates and building blocks. When target molecules fall entirely outside the accessible chemical space, ML-based methods may theoretically have advantages due to their unconstrained learned representations. However, comprehensive real-world benchmarking from USPTO and FDA-approved drugs, demonstrates that SynTwins consistently achieves superior structural similarity while guaranteeing synthesizability. This suggests that explicit pathway construction provides valuable inductive bias that outweighs theoretical flexibility in practical analog generation. Furthermore, SynTwins' algorithmic transparency enables not only a clear understanding of success and failure modes but also actionable insights, where chemists can inspect the retrosynthetic decomposition to identify exactly which building blocks or reaction templates are missing, enabling strategic and targeted expansion of chemistry resources to improve coverage systematically. This interpretability and actionability represent critical advantages for iterative discovery workflows where experimental validation and continuous improvement are essential. We anticipate that SynTwins will contribute to more efficient molecular design by facilitating the synthesis of viable molecular analogs using readily available reactions and building blocks.

## Notes and references